\documentclass[aps,prc,preprint,superscriptaddress]{revtex4-1}

\usepackage{graphics}
\usepackage{dcolumn}
\usepackage{bm}
\usepackage{floatrow}
\usepackage{booktabs}
\usepackage{multirow}
\usepackage{siunitx}
\usepackage{longtable}

\begin{document}


\title{$g_{9/2}$ neutron strength in the $N=29$ isotones and the $^{52}$Cr($d,p$)$^{53}$Cr reaction}



\author{L. A. Riley}
\affiliation{Department of Physics and Astronomy, Ursinus College,
  Collegeville, PA 19426, USA}

\author{D. T. Simms} \affiliation{Department of Physics and Astronomy, Ursinus College,
  Collegeville, PA 19426, USA}

\author{L. T. Baby} \affiliation{Department of Physics, Florida
  State University, Tallahassee, FL 32306, USA}

\author{A. L. Conley} \affiliation{Department of Physics, Florida
  State University, Tallahassee, FL 32306, USA}

\author{P. D. Cottle} \affiliation{Department of Physics, Florida
  State University, Tallahassee, FL 32306, USA}

\author{J. Esparza} \affiliation{Department of Physics, Florida
  State University, Tallahassee, FL 32306, USA}

\author{K. Hanselman} \affiliation{Department of Physics, Florida
  State University, Tallahassee, FL 32306, USA}

\author{I. C. S. Hay} \affiliation{Department of Physics, Florida State University, Tallahassee, FL  32306, USA}

\author{M. Heinze} \affiliation{Department of Physics and Astronomy, Ursinus College,
  Collegeville, PA 19426, USA}

\author{B. Kelly} \affiliation{Department of Physics, Florida
  State University, Tallahassee, FL 32306, USA}

\author{K. W. Kemper} \affiliation{Department of Physics, Florida
  State University, Tallahassee, FL 32306, USA}

\author{G. W. McCann} \affiliation{Department of Physics, Florida
  State University, Tallahassee, FL 32306, USA}

\author{R. Renom} \affiliation{Department of Physics, Florida
  State University, Tallahassee, FL 32306, USA}

\author{M. Spieker} \affiliation{Department of Physics, Florida
  State University, Tallahassee, FL 32306, USA}

\author{I. Wiedenh\"over} \affiliation{Department of Physics, Florida
  State University, Tallahassee, FL 32306, USA}

\date{\today}

\begin{abstract}

We performed a measurement of the $^{52}$Cr$(d,p)^{53}$Cr reaction at 16 MeV using the Florida State University Super-Enge Split-Pole Spectrograph (SE-SPS) and observed 26 states.  While all of the states observed here had been seen in previous $(d,p)$ experiments, we changed five $L$ assignments from those reported previously and determined $L$ values for nine states that had not had such assignments made previously.

The $g_{9/2}$ neutron strength observed in $^{53}$Cr in the present work and in the $N=29$ isotones $^{49}$Ca, $^{51}$Ti, and $^{55}$Fe via $(d,p)$ reactions is much smaller than the sum rule for this strength.  Most of the observed $L=4$ strength in these nuclei is located in states near 4 MeV excitation energy.  The remaining 
$g_{9/2}$ strength may be located in the continuum or may be fragmented among many bound states.  A covariant density functional theory calculation provides support for the hypothesis that the $g_{9/2}$ neutron orbit is unbound in $^{53}$Cr.  The ($\alpha,^3$He) reaction may provide a more sensitive probe for the missing $g_{9/2}$ neutron strength.  In addition, particle-$\gamma$ coincidence experiments may help resolve some remaining questions in this nucleus.   

\end{abstract}


\maketitle


\section{Introduction}

As Maria Goeppert Mayer pointed out in 1949 \cite{Ma49}, in nuclear shell structure the $1g_{9/2}$ orbit is the lowest-lying ``intruder'' orbit that is pushed down from its spin-orbit partner by the spin-orbit force into the next lower major shell, forming the $fpg$ shell.  The determination of the energy of the $1g_{9/2}$ neutron orbit is particularly important because of the role this orbit plays in the island of inversion phenomenon that occurs in isotopes near $^{60}$Cr (see Ref. \cite{Ga21} and references therein).  In this island of inversion, pairs of neutrons are promoted from $fp$ orbits into the $g_{9/2}$ orbit, producing well-deformed shapes.

The nuclei in which it is most straightforward to determine the single neutron energies are the isotopes that have one neutron added to a closed shell (or one neutron subtracted from a closed shell).  Nuclear reactions that deposit a single neutron onto a target with a closed neutron shell (or remove one neutron from a closed neutron shell) provide information on the energies of the single neutron orbits, even when the single-particle strength of these orbits is fragmented among several states in which the single neutron configurations mix with other nuclear excitations.  By determining the single neutron strengths (or hole strengths) in these fragments, we can calculate the single neutron energy (or single neutron hole energy) as the centroid of the observed strength.

Here we present a new measurement of $1g_{9/2}$ neutron strength in the $N=29$ isotope $^{53}$Cr via 
the $^{52}$Cr$(d,p)^{53}$Cr reaction and compare this new experimental result with recent $(d,p)$ results on the $N=29$ isotones $^{51}$Ti and $^{55}$Fe.  In these three nuclei, the sums of the $1g_{9/2}$ spectroscopic factors of the observed states are smaller than 0.5.  It is possible that much of the $1g_{9/2}$ strength may be located above the single nucleon separation thresholds.  This possibility is supported by a covariant functional theory calculation.  Another possibility is that the $g_{9/2}$ strength is so fragmented that it is difficult to observe all of the fragments.  The fragmentation of the $1g_{9/2}$ neutron strength in these isotones results at least in part from mixing with $J^{\pi}=9/2^+$ states that occur because of the coupling of the octupole excitation in the core to the $2p_{3/2}$ ground states of the $N=29$ isotones.

\section{Experimental details and results}

A deuteron beam, produced by a source of negative ions by cesium sputtering (SNICS) with a deuterated titanium cone, was accelerated to an energy of 16 MeV by the 9 MV Super FN Tandem Van de Graaff Accelerator at the John D. Fox Superconducting Accelerator Laboratory at Florida State University. The beam was delivered to a natural Cr target of thickness 300 $\mu$g/cm$^2$ on a 20 $\mu$g/cm$^2$ carbon backing that was mounted in the target chamber of the Super-Enge Split-Pole Spectrograph.  The natural abundance of $^{52}$Cr is 84\%. The spectrograph, which accepted a solid angle of 4.6 msr, was rotated from scattering angles of 15$^\circ$ to 50$^\circ$ at increments of 5$^\circ$ to measure angular distributions of protons from the $^{52}$Cr$(d,p)^{53}$Cr reaction.  Further details of the experimental setup are described in Ref.~\cite{Ri21}. 



\begin{figure*}[th]
  \begin{center}
    \scalebox{1.2}{
      \includegraphics{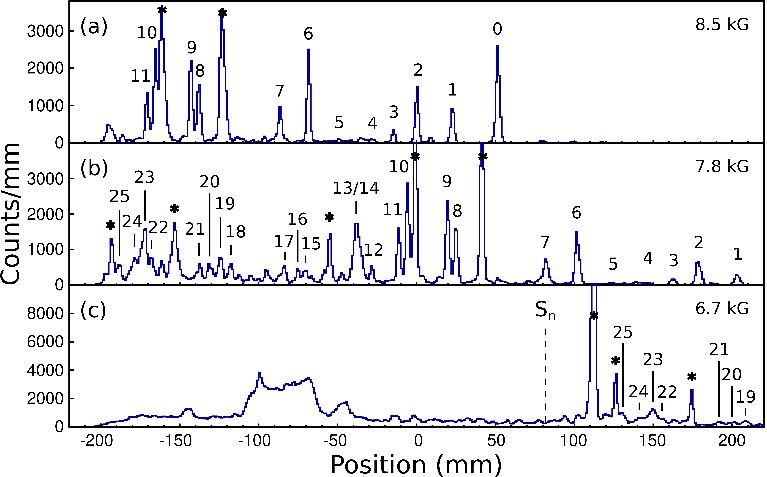}
    }
    \caption{\label{fig:spectrum} (Color online) Proton momentum spectra at a laboratory angle of 25$^\circ$ for the three magnetic-field settings used in the spectrograph for this experiment. Peaks from $^{53}$Cr are labeled with the numbers listed in Table~\ref{tab:states}.  Contaminant peaks are labeled with asterisks. The spectra are shown as a function of position in the focal plane detector.} 
  \end{center}
\end{figure*}

Proton momentum spectra collected at a scattering angle of 25$^\circ$ and with the three spectrograph magnetic field settings used in this experiment are shown in Figure~\ref{fig:spectrum}.

\begin{center}
\begin{longtable}{cccccccc}
  \caption{\label{tab:states} Excitation energies from the present work and Ref. \cite{NNDC53}, angular-momentum transfer, and $J^\pi$ assignments, single-neutron orbits used for the \textsc{fresco} \cite{Tho88} analysis, and the spectroscopic factors for states of $^{53}$Cr populated in the present work.  Established $J^\pi$ assignments are from Ref.~\cite{NNDC53}. Tentative $J^\pi$ assignments based on $L$ values determined in the present work are discussed in the text. When more than one possible orbit is given for a state, the spectroscopic factors assuming both orbits are shown.}\\

\hline

Label & $E_x$ (keV)    & $E_x$ (keV)             & \textit{L} & J$^\pi$ & orbit & S & Comments \\
         &  Present Work & Ref. \cite{NNDC53} &                 &              &          &   &                   \\
\hline
\endfirsthead

\multicolumn{7}{c}

{{\bfseries \tablename\ \thetable{} --- continued from previous page}} \\
\hline
Label & $E_x$ (keV)   & $E_x$ (keV)             & \textit{L} & J$^\pi$ & orbit & S & Comments \\
         & Present Work & Ref. \cite{NNDC53} &                 &              &         &    &                   \\
\hline
\endhead

\hline \multicolumn{7}{r}{{Continued on next page}} \\ \hline
\endfoot

\hline\hline
\endlastfoot

        0  & 0(3) & 0                                & 1 & $\frac{3}{2}^-$ & $2p_{3/2}$ & 0.33(2) & \\
        1  & 564(2)     & 564.03(4)     & 1 & $\frac{1}{2}^-$ & $2p_{1/2}$ & 0.21(2) & \\
        2  & 1006(2)   & 1006.27(5)   & 3 & $\frac{5}{2}^-$ & $1f_{5/2}$ & 0.21(1) & \\
        3  & 1289(2)   & 1289.52(7)   & 3  & $\frac{7}{2}^-$ & $1f_{7/2}$  & 0.032(3) &   \\
        4  & 1549(11)   & 1536.62(7)   & 3  & $\frac{7}{2}^-$ & $1f_{7/2}$ & 0.008(1) & \\
        5  & 1949(12)   & 1973.66(11) &  1  & $\frac{1}{2}^-$ & $2p_{1/2}$ & 0.110(24)  &  Ref. \cite{NNDC53} reports no $L$ assignment.  \\
            &                &     &   & $\frac{3}{2}^-$ & $2p_{3/2}$ & 0.055(12)  & \\
        6  & 2317(4)   & 2320.71(21) & 1  & $\frac{3}{2}^-$ & $2p_{3/2}$ & 0.15(1) &  \\ 
        7  & 2664(6)   & 2656.5(3)     & 3 & $\frac{5}{2}^-$  & $1f_{5/2}$ & 0.11(1) &  \\
        8  & 3619(9)   & 3616.51(18) & 1 & $\frac{1}{2}^-$  & $2p_{1/2}$ & 0.20(2) & \\
        9  & 3712(10) & 3706.5(15)   & 4 & $\frac{9}{2}^+$ & $1g_{9/2}$ & 0.22(1) & \\
        10 & 4170(11)  & 4135.1(6)     & 2 & $\frac{5}{2}^+$  & $2d_{5/2}$ & 0.054(4) & Ref. \cite{NNDC53} reports $J^{\pi}=5/2^+,3/2^+$ \\
        11 & 4268(10) & 4230.5(7)      & 2 & $\frac{5}{2}^+$ & $2d_{5/2}$ & 0.027(2) & Ref. \cite{NNDC53} reports $J^{\pi}=5/2^+,3/2^+$ \\
        12 & 4562(10) & 4551(10)       & 2 & $\frac{5}{2}^+$ & $2d_{5/2}$ & 0.011(1) & Ref. \cite{NNDC53} reports no $L$ assignment. \\
        13 & 4683(10) & 4690(7)        & 1 & $\frac{1}{2}^-$ & $2p_{1/2}$ & 0.10(2) & Ref.  \cite{NNDC53} reports $J^{\pi}=1/2^+$ \\
             &               &    &    & $\frac{3}{2}^-$ & $2p_{3/2}$ & 0.050(10) & \\
        14 & 4740(10) & 4745(7)        & 3 & $\frac{5}{2}^-$ & $1f_{5/2}$ & 0.15(2) & Ref. \cite{NNDC53} reports no $L$ assignment. \\ 
        15 & 5306(10) & 5310(10)      & 3 & $\frac{5}{2}^-$ & $1f_{5/2}$ & 0.026(4) & Ref. \cite{NNDC53} reports no $L$ assignment.  \\
        16 & 5379(10) & 5397(10)      & 3 & $\frac{5}{2}^-$ & $1f_{5/2}$ & 0.020(4) & Ref.  \cite{NNDC53} reports $J^{\pi}=1/2^-,3/2^-$. \\
        17 & 5529(10) & 5557(10)      & 1  & $\frac{1}{2}^-$ & $2p_{1/2}$ & 0.043(6) &  \\
             &               &    & 1  & $\frac{3}{2}^-$ & $2p_{3/2}$ & 0.022(3) & \\
        18 & 6123(10) & 6114(10)      & 3 & $\frac{5}{2}^-$ & $1f_{5/2}$ & 0.031(5) & Ref. \cite{NNDC53} reports no $L$ assignment. \\
        19 & 6230(10) & 6231(10)      & 4 & $\frac{9}{2}^+$ & $1g_{9/2}$ & 0.036(2) & Ref. \cite{NNDC53} reports $J^{\pi}=(1/2^+)$. \\
        20 & 6342(10) & 6335(10)      & 3 & $\frac{5}{2}^-$ & $1f_{5/2}$ & 0.024(2) & Ref. \cite{NNDC53} reports no $L$ assignment. \\
        21 & 6460(10) & 6460(10)      & 1 & $\frac{1}{2}^-$ & $2p_{1/2}$ & 0.044(3) &  Ref. \cite{NNDC53} reports no $L$ assignment \\
             &               &    & 1  & $\frac{3}{2}^-$ & $2p_{3/2}$ & 0.022(2) & \\
        22 & 6961(10)   & 6961(10)      & 1 & $\frac{1}{2}^-$ & $2p_{1/2}$ & 0.032(8) &  Ref. \cite{NNDC53} reports $J^{\pi}=1/2^+$.  \\
             &               &    & 1  & $\frac{3}{2}^-$ & $2p_{3/2}$ & 0.016(4) & \\
        23 & 7045(12)  & 7056(10) & 3 & $\frac{5}{2}^-$ & $1f_{5/2}$ & 0.058(4) &  Ref. \cite{NNDC53} reports no $L$ assignment \\
        24 & 7165(10)  & 7167(10) & 3 & $\frac{5}{2}^-$ & $1f_{5/2}$ & 0.022(3) & Ref. \cite{NNDC53} reports $J^{\pi}=1/2^+$. \\
        25 & 7329(10)  & 7321(10) & 2 & $\frac{5}{2}^+$ & $2d_{5/2}$ & 0.018(1) & Ref. \cite{NNDC53} reports no $L$ assignment  \\
              
\end{longtable}
\end{center}

\begin{figure*}[h]
  \begin{center}
    \scalebox{0.65}{
      \includegraphics{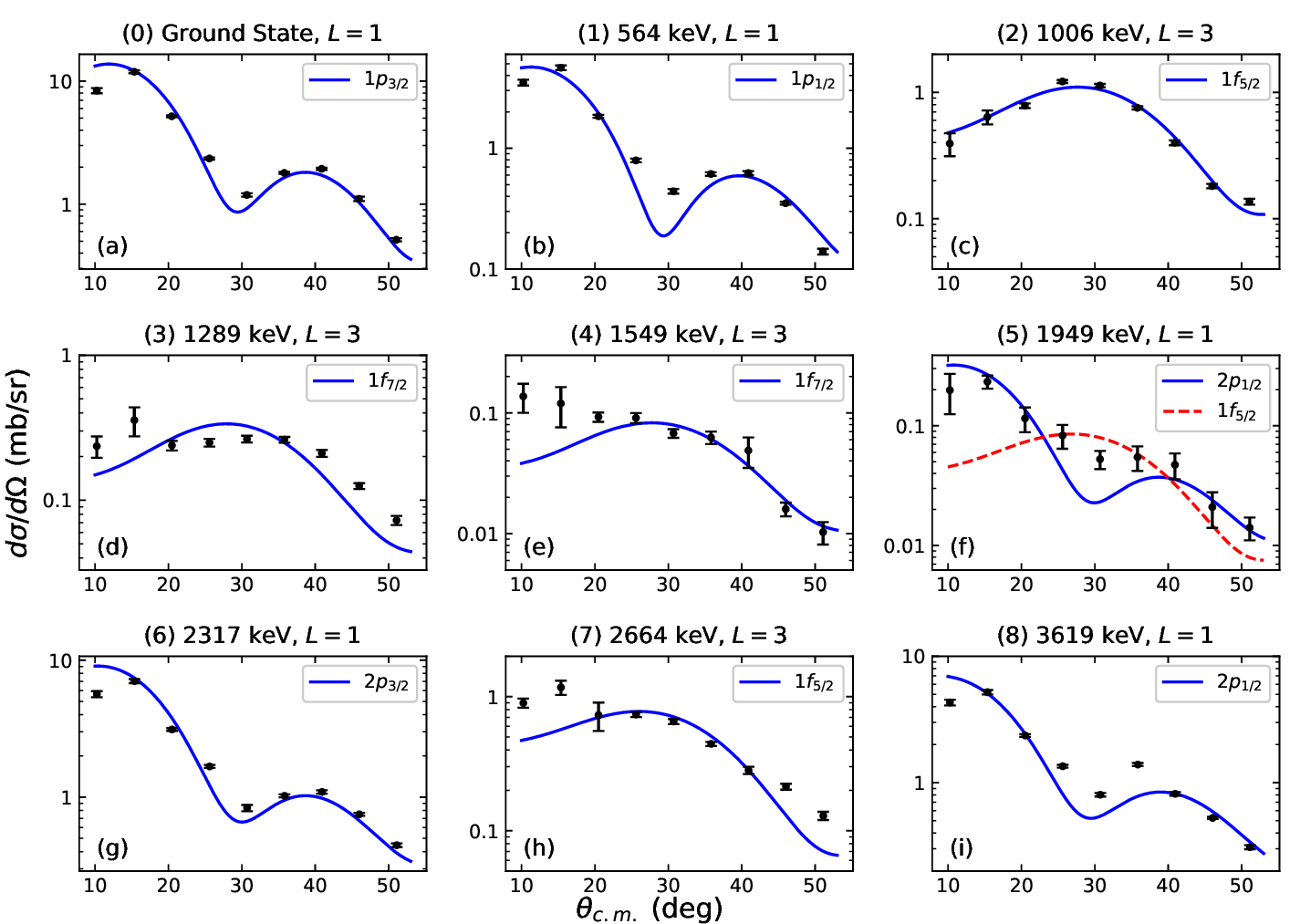}
    }
    \caption{\label{fig:angdist1} (Color online) Measured proton angular distributions from the $^{52}$Cr$(d,p)^{53}$Cr reaction compared with \textsc{fresco} \cite{Tho88} calculations described in the text.  Panels (a) to (i) correspond to the states 0-8 in Table~\ref{tab:states}.}
    \end{center}
\end{figure*}

\begin{figure*}[h]
  \begin{center}
    \scalebox{0.65}{
      \includegraphics{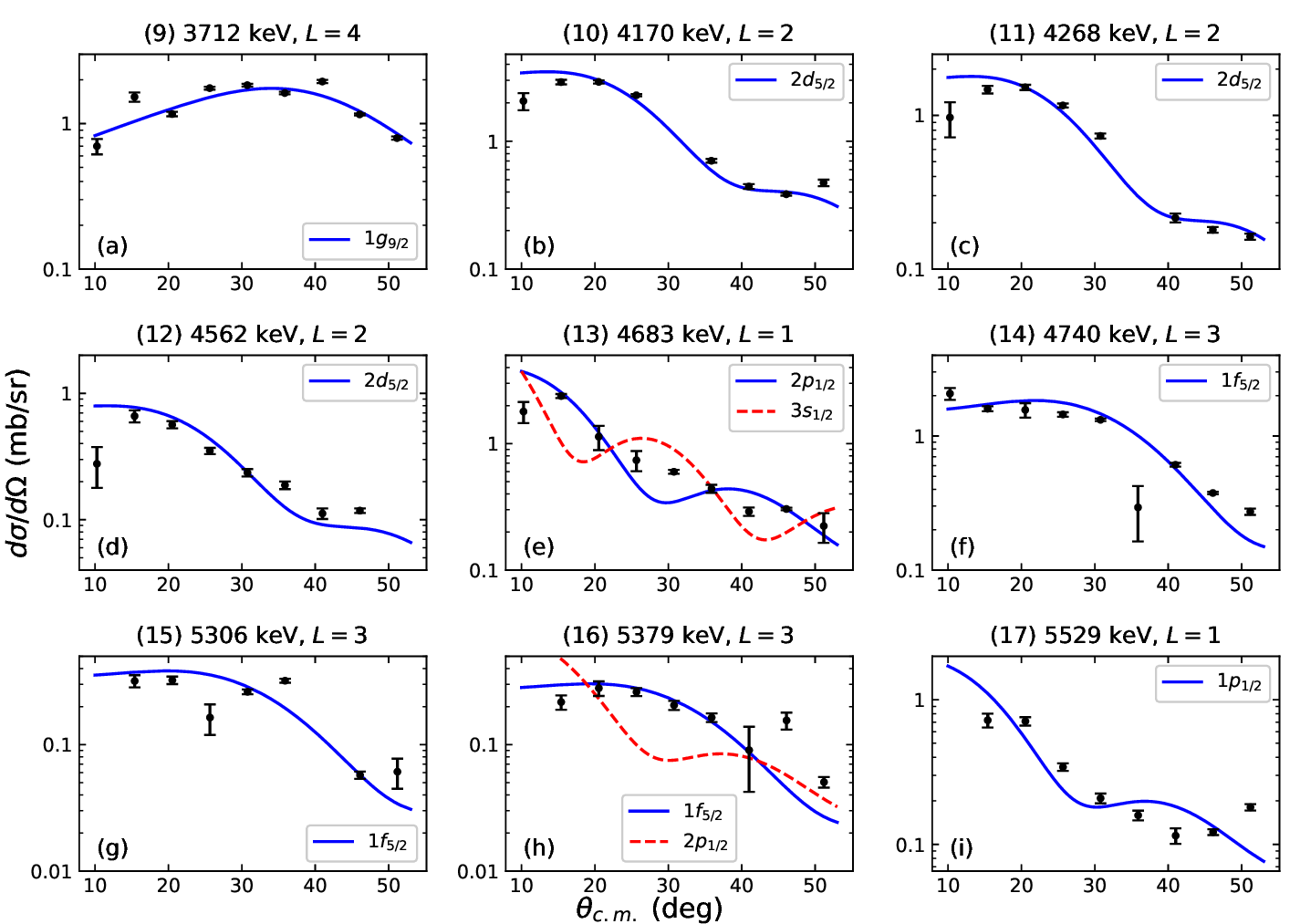}
    }
    \caption{\label{fig:angdist2} (Color online) Measured proton angular distributions from the $^{52}$Cr$(d,p)^{53}$Cr reaction compared with \textsc{fresco} \cite{Tho88} calculations described in the text.   Panels (a) to (i) correspond to the states 9-17 in Table~\ref{tab:states}.}
    \end{center}
\end{figure*}

\begin{figure*}[h]
  \begin{center}
    \scalebox{0.65}{
      \includegraphics{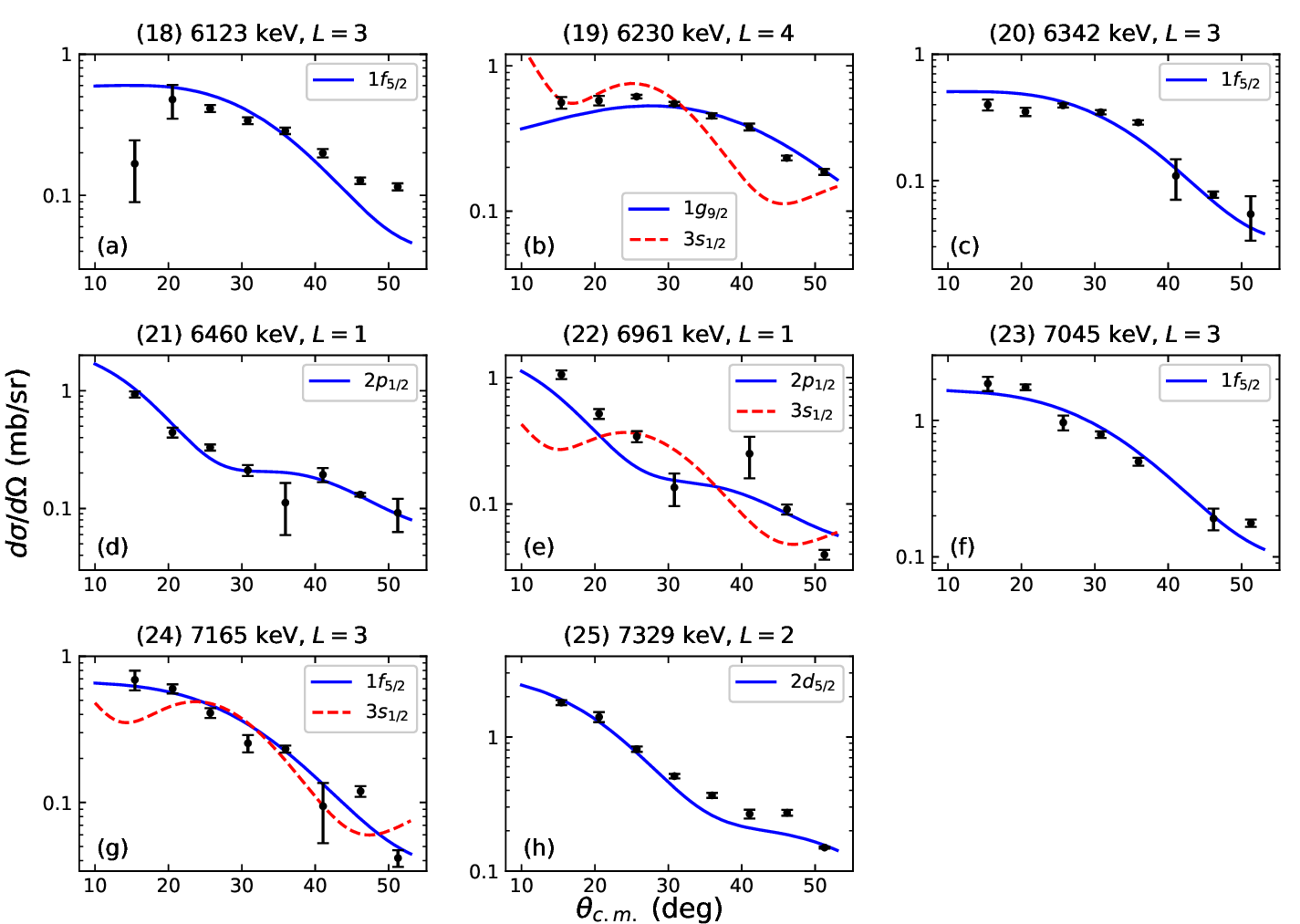}
    }
    \caption{\label{fig:angdist3} (Color online) Measured proton angular distributions from the $^{52}$Cr$(d,p)^{53}$Cr reaction compared with \textsc{fresco} \cite{Tho88} calculations described in the text.   Panels (a) to (h) correspond to the states 18-25 in Table~\ref{tab:states}.}
    \end{center}
\end{figure*}

\begin{table*}
  \caption{\label{tab:omps} Optical potential parameters used in \textsc{fresco} \cite{Tho88} calculations in the present work determined using Refs. \cite{Joh70} and \cite{Wal76} as described in the text.}

  \begin{tabular}{ccccccccccccccc}\hline\hline
             & $V_V$ & $r_V$ & $a_V$& $W_V$ & $r_W$ & $a_W$ & $W_D$ & $r_D$ & $a_D$ & $V_{so}$ & $W_{so}$ & $r_{so}$ & $a_{so}$ & $r_C$  \\
             & (MeV) & (fm) & (fm)  & (MeV) & (fm)  & (fm)  & (MeV)& (fm)  & (fm)  & (MeV)   & (MeV)    & (fm)     & (fm)    & (fm)   \\\hline
    d+$^{52}$Cr & 104.3 & 1.195 & 0.702 & 1.23  & 1.197  & 0.702 & 14.98 & 1.283  & 0.583 & 11.31    & -0.13   & 1.013     & 0.621   & 1.25   \\
    p+$^{53}$Cr &  46.4 & 1.196 & 0.670 & 1.30  & 1.197  & 0.670 &  6.88 & 1.283  & 0.553 &  5.48    & -0.08   & 1.013     & 0.590   & 1.25   \\\hline\hline
  \end{tabular}
\end{table*}

The magnetic rigidity spectrum measured at each scattering angle was fit using a linear combination of Gaussian functions with a cubic background. The proton yields corresponding to each state in \textsuperscript{53}Cr were used to produce the measured proton angular distributions shown in Figs.~\ref{fig:angdist1}-\ref{fig:angdist3}.  The absolute cross sections were determined to be accurate to an uncertainty of $15\%$, with contributions from uncertainties in charge integration, target thickness and solid angle.

The B$\rho$ calibration (which gives the energy calibration) is based on adopted energies from Ref. \cite{NNDC53}. The uncertainties are statistical—from both the peak positions from the fit and the propagated uncertainties in the calibration parameters, except in cases in which this results in a smaller uncertainty than that given in Ref. \cite{NNDC53}. In those cases, we report the uncertainty from Ref. \cite{NNDC53}.

To extract spectroscopic factors from the present angular distributions, calculations that use the adiabatic approach for generating the entrance channel deuteron optical potentials (as developed by Johnson and Soper \cite{Joh70}) were used.  The potential was produced using the formulation of Wales and Johnson \cite{Wal76}.  Its use takes into account the possibility of deuteron breakup and has been shown to provide a more consistent analysis as a function of bombarding energy \cite{De05} as well as across a large number of $(d,p)$ and $(p,d)$ transfer reactions on $Z=3-24$ target nuclei \cite{Le07}.  The proton-neutron and neutron-nucleus global optical potential parameters of Koning and Delaroche \cite{Kon03} were used to produce the deuteron potential as well as the proton-nucleus optical potential parameters needed for the exit channel of the $(d,p)$ transfer calculations, in keeping with the nomenclature of Ref. \cite{De05}.  The angular-momentum transfer and spectroscopic factors found in Table~\ref{tab:states} were determined by scaling these calculations, made with the \textsc{fresco} code \cite{Tho88}, to the proton angular distributions. Optical potential parameters are listed in Table~\ref{tab:omps}.  The overlaps between 
$^{53}$Cr and $^{52}$Cr$+n$ were calculated using binding potentials of Woods-Saxon form whose depth was varied to reproduce the given state's binding energy with geometry parameters of $r_0=1.25$ fm and $a_0=0.65$ fm and a Thomas spin-orbit term of strength $V_{so}=6$ MeV that was not varied.


We observed 26 states in $^{53}$Cr, all of which had been previously observed in $^{52}$Cr$(d,p)$ measurements \cite{NNDC53}.  However in five of these states, the transferred angular momentum $L$ determined here is different from that given in Ref. \cite{NNDC53}.  For the 4683 keV state, Ref. \cite{NNDC53} reported $J^{\pi}=1/2^+$, corresponding to $L=0$.  We determined that the 4683 keV state is populated via $L=1$ transfer instead by comparing the chi-square value of 24.2 for the best $L=1$ fit to the chi-square value of 83.1 to the best $L=0$ fit.  Similarly, we changed:  the $L$ assignment for the 5379 keV state to $L=3$ (chi-square of 7.0) from the $L=1$ value given in Ref. \cite{NNDC53} (chi-square of 34.7); the assignment for the 6230 keV state to $L=4$ (chi-square of 8.4) from the $L=0$ value given in Ref. \cite{NNDC53} (chi-square of 58.1); the assignment for the 6961 keV state to $L=1$ (chi-square of 13.2) from the $L=0$ value given in Ref. \cite{NNDC53} (chi-square of 23.6); and, the assignment for the 7165 keV state to $L=3$ (chi-square of 4.0) from the $L=0$ value given in Ref. \cite{NNDC53} (chi-square of 12.3). 

In another nine states, we made $L$ assignments for the first time.  Of these nine states, the most difficult to assign was the 1949 keV state.  As shown in Fig.~\ref{fig:angdist1}, we performed best fits for $L=1$ and $L=3$ to the data.  A comparison of the chi-square values for $L=1$ (2.6) and $L=3$ (6.5) favored the $L=1$ assignment.

Distinguishing between spin-orbit partners like $p_{3/2}$-$p_{1/2}$ and $f_{7/2}$-$f_{5/2}$ with the $(d,p)$ reaction generally requires the measurement of analyzing powers with a polarized deuteron beam, which was not available for the present experiment.  Therefore, unless there is other experimental evidence available for $L=1$ states to distinguish between $J^{\pi}=3/2^-$ and $1/2^-$ assignments, we list both possibilities (and spectroscopic factors for both possibilities) in Table~\ref{tab:states}.  We approached $L=3$ states differently because the $f_{7/2}$ orbit lies below the $N=28$ shell closure. Aside from the 1289 and 1549 keV states, we assumed that states for which angular distributions were best fit with $L=3$ were $J^{\pi}=5/2^-$ states (corresponding to the $f_{5/2}$ neutron orbit).

Only two of the states observed here have $L=4$, corresponding to neutron transfer into the $g_{9/2}$ neutron orbit.  The distribution of $g_{9/2}$ strength in the $N=29$ isotones $^{49}$Ca, $^{51}$Ti, $^{53}$Cr, and $^{55}$Fe is compared with that of the $f_{5/2}$ strength in Fig.~\ref{fig:n29_g4.5} and discussed in the next section.

\section{Discussion}

\begin{figure}[h]
  \begin{center}
    \scalebox{0.65}{
      \includegraphics{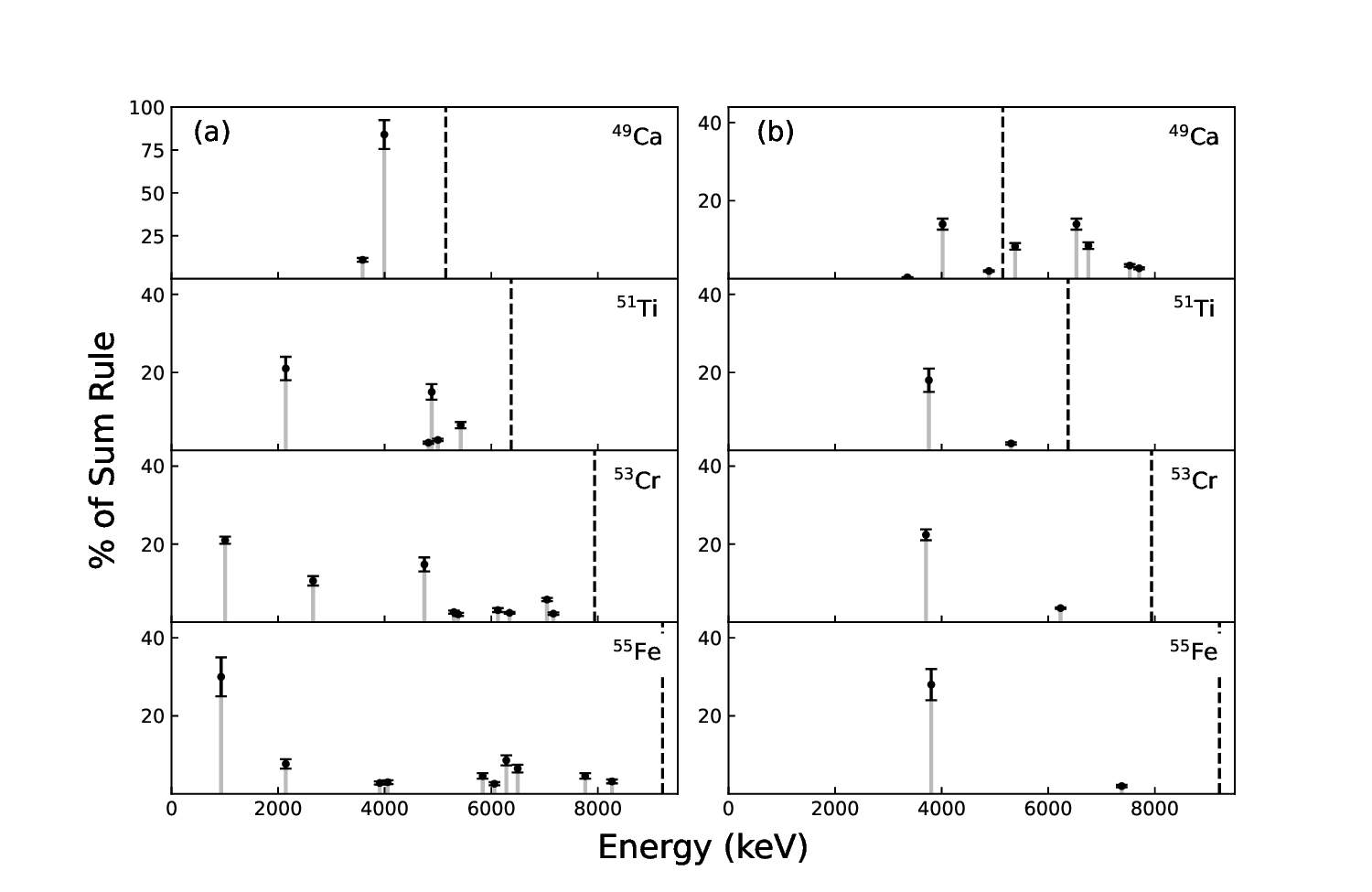}
    }
    \caption{\label{fig:n29_g4.5} The $f_{5/2}$ (panel a) and $g_{9/2}$ (panel b) strength distributions observed in the $N=29$ even-Z isotones from Ca to Fe.  A spectroscopic factor of 1 would correspond to 100\% of the sum-rule strength.  The dashed lines show the particle decay thresholds, which are the single neutron separation energies in $^{49}$Ca, $^{51}$Ti and $^{53}$Cr and the single proton separation energy in $^{55}$Fe.  The data for $^{53}$Cr are from the present work.  Data for $^{49}$Ca are from Ref. \cite{Uo94}; for $^{51}$Ti from Ref. \cite{Ri21}; and for $^{55}$Fe from 
\cite{Ri23}.  Single nucleon separation energies are from Refs. \cite{NNDC49,NNDC51,NNDC53,NNDC55}.} 
    \end{center}
\end{figure}

In $(d,p)$ studies of the even-Z $N=29$ isotones $^{49}$Ca \cite{Uo94}, $^{51}$Ti \cite{Ri21} and $^{55}$Fe \cite{Ri23}, the total spectroscopic strengths observed for the $g_{9/2}$ neutron orbit are much smaller than the strengths observed for the $f_{5/2}$ neutron orbit.  While distinguishing between $p_{3/2}$ and $p_{1/2}$ states can be difficult without analyzing power data from reactions with polarized deuteron beams, nearly all of the $L=3$ strength observed in $(d,p)$ reactions in these nuclei can be attributed to the $f_{5/2}$ orbit.  Therefore, comparing the observed $g_{9/2}$ strength with that of the $f_{5/2}$ neutron orbit is the best way of determining whether the $g_{9/2}$ strength is anomalously small.  

In the $^{48}$Ca$(d,p)^{49}$Ca study at 56 MeV by Uozumi \textit{et al.} \cite{Uo94}, the sum of the spectroscopic factors for the observed $f_{5/2}$ states is 0.97, while the sum of the $g_{9/2}$ spectroscopic strengths is 0.53.  Incidentally, Uozumi \textit{et al.} used a polarized deuteron beam so they were able to distinguish between $p_{3/2}$ and $p_{1/2}$ neutron states.  The sum of the spectroscopic factors for the $p_{3/2}$ states Uozumi \textit{et al.} observed was 0.97, while the sum of the spectroscopic factors they obtained for $p_{1/2}$ was 1.03.

The most recent $(d,p)$ study of $^{51}$Ti was performed by Riley \textit{et al.} at 16 MeV \cite{Ri21}.  In this study, the sum of the spectroscopic factors for the $f_{5/2}$ states was 0.47(4), while the corresponding sum for the $g_{9/2}$ states was 0.20(3).  

In $^{55}$Fe, Riley \textit{et al.} \cite{Ri23} used the $(d,p)$ reaction at 16 MeV to identify several $f_{5/2}$ states that gave a summed spectroscopic factor of 0.74(6).  In the same study, the sum of spectroscopic factors for $g_{9/2}$ was 0.32(4).

In all three of the cases, the observed $g_{9/2}$ strength was less than 60\% of the $f_{5/2}$ strength.

In the present study of $^{53}$Cr, the sum of the spectroscopic factors listed for the two states listed in Table 1 that are populated via $L=4$ transfer (and which are therefore presumed to be $g_{9/2}$ neutron states) is 0.26(1).  However, the sum of the spectroscopic factors for the $f_{5/2}$ states measured in the present experiment is 0.57(3).  
In $^{53}$Cr, as in $^{49}$Ca, $^{51}$Ti and $^{55}$Fe, the observed $g_{9/2}$ strength is much smaller than the observed $f_{5/2}$ strength. 

The distributions of $f_{5/2}$ and $g_{9/2}$ strength in these four nuclei are summarized in Fig. \ref{fig:n29_g4.5}. 

It is clear that in all four of these $N=29$ isotones, the $g_{9/2}$ neutron strength is fragmenting by mixing with other $J^{\pi}=9/2^+$ states and that this is resulting in a significant share of the $g_{9/2}$ strength being concentrated in a state near 4.0 MeV.  One way to produce a $9/2^+$ state in these $N=29$ isotones is to couple the $p_{3/2}$ neutron, which is the lowest valence neutron orbit in these isotones and which sets the $3/2^-$ ground state $J^{\pi}$ values in all four of them, to the low energy octupole state in the $N=28$ core nucleus.  In $^{49}$Ca, Montanari \textit{et al.} \cite{Mo12} demonstrated that the $9/2^+$ state at 4017.5 keV has a large octupole component.  They populated $^{49}$Ca via a single-neutron transfer reactions with a $^{48}$Ca beam impinging on $^{64}$Ni and $^{208}$Pb targets and used a large array of high-resolution $\gamma$-ray detectors to measure lifetimes with the differential recoil distance Doppler-shift method.  They were able to determine that the reduced matrix element $B(E3)$ for the decay of the 
4017.5 keV $9/2^+$ state to
the $3/2^-$ ground state is 8(2) W.u.  This result overlaps with the value of 8.4 W.u. (+4.3, -3.5) given in Ref. \cite{Ch22} for the transition from the lowest $3_1^-$ state in the core nucleus $^{48}$Ca (which is located at 4507 keV) to the ground state.  But in addition, Uozumi \textit{et al.} \cite{Uo94} determined that the 4017.5 keV state in $^{49}$Ca has a $g_{9/2}$ neutron spectroscopic factor of 0.14.  So clearly this state has a significant $g_{9/2}$ single neutron component as well.

The situations in $^{51}$Ti, $^{53}$Cr and $^{55}$Fe appear to be similar to that in $^{49}$Ca.  In $^{51}$Ti, there is a $9/2^+$ state at 3771 keV that has a $g_{9/2}$ spectroscopic factor of 0.18(3) \cite{Ri21}.  In the $^{50}$Ti core nucleus, the $3^-$ state that appears to be the strongest low energy octupole state occurs at 4410 keV \cite{Ch19}.  The lowest 
$9/2^+$ state in $^{53}$Cr, which occurs at 3706 keV and has a $g_{9/2}$ spectroscopic factor of 0.22(3), can be compared in energy to the $3_1^-$ state in $^{52}$Cr, which occurs at 4470 keV \cite{Do15}.  In $^{55}$Fe, the lowest $9/2^+$ state is found at 3804 keV and has a $g_{9/2}$ spectroscopic factor of 0.28(4) \cite{Ri23}.  The $3_1^-$ state in the core 
nucleus $^ {54}$Fe occurs at 4782 keV \cite{Do14}.

Mixing between a $g_{9/2}$ single neutron state and a $p_{3/2} \bigotimes 3_1^-$ state that occurs at a lower energy than the unperturbed $g_{9/2}$ neutron state would certainly result in what we see experimentally in $^{49}$Ca and what we likely have in $^{51}$Ti, $^{53}$Cr and $^{55}$Fe as well---a $9/2^+$ state that has a somewhat collective $B(E3)$ value for decay to the ground state and a $g_{9/2}$ spectroscopic factor that is significant but much smaller than 1.0.  But this two-state mixing scenario would also result in another state at higher energy than the unperturbed $g_{9/2}$ single neutron state that carries most of the $g_{9/2}$ strength.  At present, there is no evidence for such a state or even a high-lying concentration of $L=4$ strength in the four $N=29$ isotones being discussed here.     

The present $^{53}$Cr experiment and the recent experiments on $^{51}$Ti \cite{Ri21} and $^{55}$Fe \cite{Ri23} only searched for states up to the particle thresholds (6372 keV in $^{51}$Ti, 7939 keV in $^{53}$Cr and 9213 keV 
in $^{55}$Fe \cite{NNDC51,NNDC53,NNDC55}).  Therefore, it is at least possible that the bulk of the $g_{9/2}$ strength is in the continuum.

The possibility that the bulk of the $g_{9/2}$ neutron strength is in the continuum is given credibility by the results
of a calculation performed in the framework of covariant density-functional theory.  This calculation of the binding energies of the $p_{3/2}$, $p_{1/2}$, $f_{5/2}$ and $g_{9/2}$ neutron orbits in
$^{48}$Ca, $^{50}$Ti, $^{52}$Cr and $^{54}$Fe uses the covariant energy density functional FSUGarnet \cite{che15} and is described in detail in Ref. \cite{Ri21}.

Table~\ref{tab:binding} shows that the calculations for the  $p_{3/2}$, $p_{1/2}$ and $f_{5/2}$ binding energies in $^{48}$Ca, $^{50}$Ti and $^{54}$Fe are within 0.7 MeV of the experimental binding energies for these orbits in $^{49}$Ca, $^{51}$Ti and $^{55}$Ti.  That is, the calculation provides a reasonable description of the binding energies of these neutron orbits.  The same calculation predicts that the $g_{9/2}$ neutron orbit is unbound in $^{48}$Ca, $^{50}$Ti and $^{52}$Cr, and bound by only 1.4 MeV in $^{54}$Fe.     

It is also possible that the $g_{9/2}$ neutron orbit is bound and that the strength is located in bound states, but the strength is so fragmented that the present experiments do not have the sensitivity necessary to observe it.  

In either case, finding the ``missing'' $g_{9/2}$ neutron strength would require a more sensitive experimental probe than the ($d,p$) reaction with 16 MeV deuterons used in the present work and in Refs. \cite{Ri21,Ri23}.  As noted by (for example) Szwec \textit{et al.} \cite{Sz21}, single nucleon transfer reactions vary in their sensitivities to populating orbits of different $L$ values.  In the reaction studied in the present work, the difference in the angular momenta of the incoming deuteron and outgoing proton is $1.0\hbar$.  Therefore, this reaction is most sensitive to the $p_{3/2}$ and $p_{1/2}$ orbits.  In contrast, the ($\alpha,^3$He) reaction is more sensitive to orbits with higher angular momenta.  For example, the difference between the angular momenta of the incoming $\alpha$-particle and outgoing $^3$He nucleus in the $^{52}$Cr($\alpha,^3$He)$^{53}$Cr reaction at 32 MeV (an energy that is accessible at the Fox Laboratory) is $6.6\hbar$.  Consequently, this reaction would be more sensitive to neutron orbits having larger orbital angular momenta such as $g_{9/2}$.  

Detecting $\gamma$ rays in coincidence with particle detection in the SE-SPS could provide additional selectivity that would be especially helpful in reactions like the one studied here in which the spectrum of excited states is crowded.  CeBr$_3$ scintillators can provide resolution of $4\%$ or better at energies above 500 keV while providing resilience in the presence of large neutron fluxes like those present during $(d,p)$ experiments \cite{SpPC}.  Five CeBr$_3$ detectors are already available for particle-$\gamma$ coincidence experiments at the SE-SPS.

 \begin{table*}
  \caption{\label{tab:binding} Experimental binding energies in the even-Z $N=29$ isotones and theoretical binding energies for the neighboring $N=28$ isotones calculated using the covariant functional theory described in the text.  All energies in MeV.  Data taken from Refs. \cite{Uo94,Ri21,Ri23,NNDC49,NNDC51,NNDC53,NNDC55} and the present work.}
  \begin{tabular}{cccccccc}\hline\hline
             & $p_{3/2}$ & $p_{1/2}$  & $f_{5/2}$ & $p_{3/2}$  & $p_{1/2}$ & $f_{5/2}$ & $g_{9/2}$  \\
	  & expt & expt & expt & theory & theory & theory & theory \\
             \\\hline
$^{49}$Ca & 4.60(7) & 2.87(3) & 1.19(1) & 4.37  & 3.06  & 1.31 & unbound  \\
$^{51}$Ti & 5.80(15) & 4.34(22) & 2.63(10) & 5.51  & 4.21  & 3.01 & unbound  \\ 
$^{53}$Cr & 6.05(114) & 5.27(60) & 4.26(20) & 6.65  & 5.39  & 4.73 & unbound  \\
$^{55}$Fe & 8.22(11) & 6.13(22) & 5.72(18) & 7.78 & 6.58 & 6.44 & 1.42 \\   

  \\\hline\hline
  \end{tabular}
\end{table*}

\section{Conclusions}

We performed a measurement of the $^{52}$Cr($d$,$p$)$^{53}$Cr reaction at 16 MeV using the FSU SE-SPS. All 26 states we observed had been seen in previous ($d$,$p$) measurements.  However, we changed five $L$ assignments from those reported previously.  In addition, we determined $L$ values for nine states that were previously observed but for which no $L$ assignment had been made.

The $g_{9/2}$ neutron strength observed via the $(d,p)$ reaction is much less than expected in the $N=29$ isotones $^{49}$Ca, $^{51}$Ti, $^{53}$Cr and $^{55}$Fe.  Most of the observed $g_{9/2}$ strength in these nuclei is located in states near 4 MeV.  The remaining 
$g_{9/2}$ strength may be located in the continuum.  This possibility is supported by the convariant functional theory calculation presented here.  Alternatively, the $g_{9/2}$ strength may be fragmented among many bound states.  The ($\alpha,^3$He) reaction may provide a more sensitive probe for the missing $g_{9/2}$ neutron strength.  
In addition, particle-$\gamma$ coincidence experiments with CeBr$_3$ detectors may provide additional sensitivity for identifying these missing fragments.

\begin{acknowledgments}
This work was supported by the National Science Foundation through grant number PHY-2012522.  We thank Jorge Piekarewicz for the covariant density functional
theory calculations.
\end{acknowledgments}


%


\end{document}